\documentclass[11pt]{article}
\usepackage{graphicx}
  \usepackage{amsthm,amsfonts}
  \usepackage{amsmath}
  \usepackage{color}
\newcommand{\bea}   {\begin{eqnarray}}
\newcommand{\eea}   {\end{eqnarray}}

 \def\*{{\boldsymbol{*}}}

\def\be{\begin{equation}}
\def\ee{\end{equation}}

\def\a{\alpha}

\def\cQ{{\mathcal{Q}}}

\begin{document}
\renewcommand{\thefootnote}{\fnsymbol{footnote}}

\thispagestyle{empty}

\title{Higher Spin Symmetries and Deformed Schr\"odinger \\Algebra in Conformal Mechanics}

\author{Francesco Toppan\thanks{{E-mail: {\em toppan@cbpf.br}}} \quad and\quad Mauricio Valenzuela\thanks{{E-mail: {\em mauricio.valenzuela@uss.cl}}}
\\
\\
}
\maketitle
\centerline{
{\it $~^\ast$ CBPF, Rua Dr. Xavier Sigaud 150, Urca,}}
{\centerline {\it\quad
cep 22290-180, Rio de Janeiro (RJ), Brazil.}}
~\\
\centerline{
{\it $~^\dag$ Facultad de Ingenier\'ia y Tecnolog\'ia,}}
{\centerline {\it\quad
Universidad San Sebasti\'{a}n, General Lagos 1163, Valdivia 5 110693, Chile.}}
~\\

\begin{abstract}
The dynamical symmetries of $1+1$-dimensional Matrix Partial Differential Equations with a Calogero potential (with/without the presence of an extra oscillatorial De Alfaro-Fubini-Furlan, DFF, term) are investigated. The first-order invariant differential operators induce several invariant algebras and superalgebras. Besides the $sl(2)\oplus u(1)$ invariance of the Calogero Conformal Mechanics, an $osp(2|2)$ invariant superalgebra, realized by first-order and second-order differential operators, is obtained. The invariant algebras with an infinite tower of generators are given by  the universal enveloping algebra of the deformed Heisenberg algebra, which is shown to be equivalent to a deformed version of the Schr\"odinger algebra.  This vector space also gives rise to a higher spin (gravity) superalgebra.
We furthermore prove that the pure and DFF Matrix Calogero PDEs possess isomorphic dynamical symmetries, being related by a similarity transformation and a redefinition of the time variable.
~\\\end{abstract}
\vfill

\rightline{CBPF-NF-001/17
}

\newpage

\section{Introduction}

In this paper we investigate the dynamical symmetries of $1+1$-dimensional Matrix Partial Differential Equations with a Calogero
\cite{cal} potential (with/without the presence of an extra oscillatorial De Alfaro-Fubini-Furlan, DFF, \cite{dff} term).
The Matrix Partial Differential Equations correspond to Schr\"odinger Equations. The pure Calogero Hamiltonian is
${\cal N}=2$ Supersymmetric, while the DFF Hamiltonian is, as explained later, only ``softly" supersymmetric. The first-order invariant differential operators induce several invariant algebras and superalgebras. Besides the $sl(2)\oplus u(1)$ invariance of the Calogero Conformal Mechanics, an $osp(2|2)$ invariant superalgebra, realized by first-order and second-order differential operators, is obtained. The invariant (super)algebras with an infinite tower of generators are given by the universal enveloping algebra of a deformed Heisenberg algebra \cite{wig,Yang1951,Plyushchay1997, vas}. It also contains a deformed version of the Schr\"odinger algebra (in both cases the deformation stems from the presence of a Klein operator). As a vector space the universal enveloping algebra carries the ${\mathbb Z}_2$-graded higher spin superalgebra $q(2;\nu)$.\par
The higher spin superalgebra $q(2;\nu)$ was introduced in \cite{vas} and  applied to the construction of a Chern-Simons higher spin gravity. In \cite{vas} it was also shown that $q(2;\nu)$ is provided by the universal enveloping algebra of the $osp(2|2)$ superalgebra.  The $q(2;\nu)$ algebra was also employed to extend the Chern-Simons higher spin gravity model \cite{vas} with fractional spin fields \cite{bsv,bsv2}.  \par
We will show that these algebraic structures appear in the different context of non-relativistic quantum mechanics in the presence of a Calogero potential, either with or without an extra harmonic potential term.
We highlight the main features and results of our approach.\par
We use the standard tools (see, e.g., \cite{olv}) of dynamical symmetries of (matrix) PDEs. The operators under consideration are
{\em local} (matrix) differential operators, in contrast with the non-local realizations given in \cite{wig,Yang1951,Plyushchay1997} (see also \cite{ply}). \par
The construction based on an invariant PDE allows us to detect an invariant deformed Schr\"odinger algebra which has not been discussed in the existing literature. This deformation can be regarded as a deformation of the mass-central-charge of the algebra of Galileo boosts and translations by a term proportional to the Klein operator. This algebra is connected with Vasiliev's $q(2;\nu)$ superalgebra. As a vector space it is spanned by the same infinite set of differential operators which, on the other hand, are endowed with different brackets (ordinary commutators versus ${\mathbb Z}_2$-graded commutators).  We furthermore prove that the models with pure Calogero potentials and those with the extra DFF term possess isomorphic dynamical symmetry (super)algebras. This results from the fact that a triplet of Matrix PDEs close an $sl(2)$ algebra. In a given canonical form, the Matrix PDE associated with the positive root of $sl(2)$ corresponds to the 
Schr\"odinger equation of the pure Matrix Calogero system (it would have been tantamount to work with the negative root of $sl(2)$),
while a canonical form exists such that the $sl(2)$ Cartan generator is associated with the Schr\"odinger equation for the DFF Matrix Calogero system. The two canonical forms are related by a similarity transformation coupled with a redefinition of the time variable.\par
The pure Matrix Calogero Hamiltonian is an ${\cal N}=2$ Supersymmetric Quantum Mechanical System. On the other hand the DFF Matrix Calogero Hamiltonian is not supersymmetric (this is a common feature of superconformal models possessing an oscillatorial term, see \cite{hoto}); it is, nevertheless, related to supersymmetric Hamiltonians via a shift realized by a diagonal operator; following \cite{cuhoto} we can state that, in this case, the system enjoys a ``soft" supersymmetry.\par
In the last part of the paper we discuss the appearance of the higher spin superalgebra, which is now interpreted as a dynamical symmetry of the models under consideration (the higher-spin symmetry of the free non-relativistic particle case was pointed out in \cite{Valenzuela2016}).\par
The scheme of the paper is as follows. In Section {\bf 2} the Matrix Calogero PDEs (with/without the DFF oscillatorial potential term) are introduced. It is explained why, in the presence of a Calogero potential, a Matrix PDE is needed to get (deformed) Heisenberg symmetry generators. The dynamical symmetry generators of these systems are computed. In Section {\bf 3} their dynamical symmetry subalgebras are investigated. In particular the $osp(2|2)$ superalgebra (realized by first-order and second-order differential symmetry operators) is derived, as well as the deformed Heisenberg algebra. The existence of an ${\cal N}=2$ supersymmetry for the pure Matrix Calogero Hamiltonian and a ``soft" supersymmetry for the DFF Matrix Calogero Hamiltonian is pointed out. The connection between pure and DFF Matrix Calogero PDEs is presented in Section {\bf 4}.
In Section {\bf 5} it is shown that both a Deformed Schr\"odinger algebra and a higher spin superalgebra appear as dynamical symmetries of (pure and DFF) Matrix Calogero PDEs. In the Conclusions we comment our results and sketch some directions for future investigations.

\section{Matrix PDEs with Calogero potentials}

It is convenient to systematically review the arising of higher-spin (super)algebras in Calogero systems by analyzing the symmetry, realized by first-order differential operators, of the (matrix) Partial Differential Equations containing Calogero potentials. This analysis uses well-known methods.\par
To fix our conventions, if $\Omega=\Omega^\dagger$ is a hermitian (matrix) PDE,  a first-order differential operator $\Sigma$ is a symmetry operator if it satisfies the equation
\bea\label{symmetry}
\relax [\Sigma, \Omega]&=& \Phi_{\Sigma}\cdot \Omega,
\eea
 for a given matrix-valued function $\Phi_\Sigma$.\par
In our analysis we do not need to introduce symmetry operators which, unlike (\ref{symmetry}), are defined by a ``right" convention, i.e. $
\relax [\Sigma, \Omega]= \Omega \cdot \Phi'_{\Sigma}$.  One should note, on the other hand, that the symmetry operator $\Sigma$ is not necessarily hermitian (in that case its hermitian conjugate $\Sigma^\dagger$ satisfies the ``right" equation with $\Phi_{\Sigma^\dagger}'=-\Phi_\Sigma^\ast$).\par
Before investigating matrix PDEs (our results require a $2\times 2$ matrix whose entries are differential operators) we recall some basic features of the standard Partial Differential Equations.  The Schr\"odinger equation in $1+1$ dimension induces  the maximal $6$-generator Schr\"odinger algebra of invariant operators  for three choices (up to normalization and trivial shift) of the potential $V(x)$ \cite{nie1,nie2,nie3,toppr}. $V(x)$ is either constant (corresponding to the free equation), linear (giving rise to Airy's functions) or quadratic (corresponding to the harmonic oscillator). For all other choices of the potential the symmetry algebra possesses fewer generators. This is particularly the case for the Calogero potential. 
\subsection{The ordinary Calogero PDEs}
The pure Calogero PDE is defined by the operator
\bea\label{omegacal}
\Omega_{Cal}&=& i\partial_t+\frac{1}{2}\partial_x^2-\frac{g}{x^2},
\eea
where $g$ is the coupling constant.
\par
We mention that the operator (\ref{omegacal}) can be applied to describe the dynamics of a two-particle Calogero model with omitted center of mass degree of freedom.  The operator also appears in the dynamics of a test particle in the proximity of the the horizon of a Reissner-Nordstr\"om black hole, see {\cite{rnbh}.
The symmetry algebra for the Calogero potential is $sl(2)\oplus(1)$. It contains four generators. Two extra generators, corresponding to the Heisenberg subalgebra of the Schr\"odinger algebra, are no longer encountered in the presence of a non-vanishing, $g\neq 0$, Calogero potential. We have in this case, as suitably normalized symmetry generators, 
\bea\label{purecal}
c&=& 1,\nonumber\\
z_+&=& \partial_t,\nonumber\\
z_0&=& -2t\partial_t-x\partial_x-\frac{1}{2},\nonumber\\
z_-&=& -t^2\partial_t-tx\partial_x+i\frac{x^2}{2}-\frac{t}{2}.
\eea
They satisfy the non-vanishing commutators
\bea\label{sl2}
\relax [z_0,z_\pm] = \pm 2 z_\pm, &&
\relax [z_+,z_-]= z_0,
\eea
while $c$ is the identity operator. \par
The non-vanishing commutators involving $\Omega_{Cal}$ are
\bea\label{sl2omegacal}
\relax [z_0,\Omega_{Cal}] = 2\Omega_{Cal}, &&
\relax [z_-,\Omega_{Cal}] = 2t\Omega_{Cal}.
\eea
Based on (\ref{symmetry}), this means that
\bea\label{symmetryrhs}
\Phi_{z_0} =2,&&
\Phi_{z_-} = 2t.
\eea
The conveniently chosen symmetry operators
\bea\label{hersymmetry}
{\overline z}_+ = i z_+, &&
{\overline z}_0= i(z_0-c) \equiv i(z_0-1),
\eea
are hermitian.\par
One should note that the four (\ref{purecal}) symmetry operators do not depend on the Calogero coupling constant $g$.\par
In the case of the Calogero potential with the extra addition, see \cite{dff}, of the DFF quadratic term, we can introduce without loss of generality (by suitably normalizing the frequency of the harmonic oscillator), the
$\Omega_{DFF}$ operator given by
\bea\label{omegadff}
\Omega_{DFF} &=&  i\partial_t+\frac{1}{2}\partial_x^2-\frac{g}{x^2}-\frac{1}{2}x^2.
\eea
In this case as well we obtain a $sl(2)\oplus u(1)$ symmetry algebra.  Its four generators are
\bea\label{dffcal}
\widehat{c} &=& 1,\nonumber\\
\widehat{z}_+&=& \frac{1}{2} e^{-2it}(\partial_t-ix\partial_x+ix^2-\frac{i}{2}),\nonumber\\
\widehat{z}_0&=& i\partial_t,\nonumber\\
\widehat{z}_-&=& \frac{1}{2}e^{2it}(\partial_t+ix\partial_x+ix^2+\frac{i}{2}).
\eea
The non-vanishing commutators are
\bea\label{newsl2}
\relax [\widehat{z}_0,\widehat{z}_\pm] = \pm 2 {\widehat z}_\pm, &&
\relax [\widehat{z}_+,\widehat{z}_-]= \widehat{z}_0.
\eea
It is worth pointing out that, when $g\neq 0$, for no other non-constant $w(x)$, the potential $\frac{g}{x^2}+w(x)$ produces a $4$-generator symmetry algebra. In particular, for a linear $w(x)$ ($w(x)=kx$, $k\neq 0$),  we only recover a $2$-generator symmetry algebra.\par
The arising of a (generalized) Heisenberg and of higher-spin (super)algebras requires a matrix Calogero PDE.  For this purpose a $2\times 2$ matrix system is sufficient.

\subsection{The $2\times 2$ -matrix Calogero PDEs}
We  check under which conditions we can obtain off-diagonal symmetry generators for a $2\times 2$-matrix differential operator containing Calogero potentials.\par
Let us denote with $e_{ij}$ ($i,j=1,2$) the matrix with entry $+1$ at the $i$-th column and $j$-th row and $0$ otherwise. We are looking for symmetry generators of the matrix PDE defined by the diagonal operator
\bea\label{diagop}
\Omega&=& (e_{11}+e_{22})(i\partial_t+\frac{1}{2}\partial_x^2) -v_1(x)e_{11}-v_2(x)e_{22}.
\eea
Since $\Omega$ is a diagonal operator, from either (\ref{purecal}) or (\ref{dffcal}), we obtain $4+4$ diagonal symmetry generators ($4$ of them are associated with the upper diagonal element $e_{11}$; the remaining $4$ generators are associated with the lower diagonal element $e_{22}$).\par
Concerning the non-diagonal symmetry operators, they should be expressed as
\bea\label{sigmaupdown}
\Sigma_{up} &=& (f(x,t)\partial_t+g(x,t)\partial_x+h(x,t))e_{12},\nonumber\\
\Sigma_{down} &=& ({\overline f}(x,t)\partial_t+{\overline g}(x,t)\partial_x+{\overline h}(x,t))e_{21},
\eea
for appropriate functions $f(x,t), g(x,t), h(x,t), {\overline f}(x,t), {\overline g}(x,t), {\overline h}(x,t)$.\par
We assume the potentials $v_1(x), v_2(x)$ to be
\bea\label{potentials}
v_i(x) &=& \frac{a_i}{x^2}+ b_ix^2+c_i,
\eea
with $a_2\neq a_1$. Without loss of generality we can set, via similarity transformations, $c_1=c_2=0$. \par
It is easily realized that the requirement
\bea\label{osceq}
b_2&=& b_1
\eea
is implied by the existence of a non-vanishing solution $\Sigma_{up}$ ($\Sigma_{down}$) for the respective equations
\bea\label{updown}
\relax [\Sigma_{up},\Omega]= 0, &&
\relax [\Sigma_{down},\Omega]=0.
\eea
The constraint on the functions entering $\Sigma_{up}$, $\Sigma_{down}$ are derived with standard techniques. We need to make vanishing the coefficients of the operators entering the r.h.s. of equations (\ref{updown}). We get, for $\Sigma_{up}$,
\bea\label{constraints}
\partial_x\partial_t&:& f_x=0  ,\nonumber\\
\partial_x^2 &:& g_x=0 ,\nonumber\\
\partial_t&:& -i{\dot f}-\frac{1}{2}f_{xx}+(v_1-v_2)f=0\quad \Rightarrow f=0\quad for\quad v_2\neq v_1, \nonumber\\
\partial_x&:& -i{\dot g} -h_x+(v_1-v_2)=0,\nonumber\\
1&:& -i{\dot h} -\frac{1}{2}h_{xx}+(v_1-v_2)h-g{v_2}_x=0.
\eea
Similar equations are obtained for ${\overline f}(x,t), {\overline g}(x,t), {\overline h}(x,t)$ entering $\Sigma_{down}$.\par
The above system of equations tells us that, in particular, the condition 
\bea\label{a1a2condition}
a_1+a_2-a_1^2-a_2^2+2a_1a_2&=&0
\eea 
has to be fulfilled. After setting $\nu= 2(a_2-a_1)$,
it is solved by
\bea\label{a1a2nu}
a_1 = \frac{1}{8}\nu (\nu-2),&&
a_2 = \frac{1}{8}\nu(\nu+2).
\eea
We can therefore write the most general (\ref{diagop}) operator admitting off-diagonal symmetry operators. It depends on the parameter $\epsilon$, whose values are either $\epsilon=0$ for
the pure Calogero system, or $\epsilon=1$ for the the Calogero system with the extra oscillatorial term. Due to (\ref{osceq}), the oscillatorial term is the same in both  upper and lower diagonal entries. We can write, with a proper normalization, 
\bea\label{matrixop}
\Omega_\epsilon &=& (e_{11}+e_{22})(i\partial_t +\frac{1}{2}\partial_x^2-\frac{1}{2}\epsilon x^2-\frac{\nu^2}{8 x^2}) +(e_{11}-e_{22})\frac{\nu}{4 x^2}.
\eea
The operator $K$,
\bea\label{klein}
K&=& e_{11}-e_{22},
\eea
plays different roles, depending on the context. It is either the Fermion Parity Operator (in Supersymmetric Quantum Mechanics) or the Klein Operator (entering the definition of the deformed oscillators). We note that, in particular, $K^2={\mathbb I}_2$.\par
In both $\epsilon=0,1$ cases we get two upper triangular and two lower triangular symmetry operators.\par
At $\epsilon =0$ we have, up to normalization,
\bea\label{sigmas}
\Sigma_1&=& e_{12} (\partial_x +\frac{\nu}{2x}),\nonumber\\
\Sigma_2 &=& e_{12}(t\partial_x +t\frac{\nu}{2x}-ix),\nonumber\\
\Sigma_3&=& e_{21}(\partial_x-\frac{\nu}{2x}),\nonumber\\
\Sigma_4&=& e_{21}(t\partial_x-t\frac{\nu}{2x}-ix).
\eea
At $\epsilon=1$ we have, up to normalization,
\bea\label{xis}
\Xi_1&=& e_{12} e^{it}(\partial_x+\frac{\nu}{2x}+x),\nonumber\\
\Xi_2&=&e_{12}e^{-it}(\partial_x+\frac{\nu}{2x}-x),\nonumber\\
\Xi_3&=& e_{21}e^{it}(\partial_x-\frac{\nu}{2x}+x),\nonumber\\
\Xi_4&=& e_{21}e^{-it}(\partial_x-\frac{\nu}{2x}-x).
\eea
\section{Symmetry superalgebra of the pure Calogero Matrix system}

For the $2\times 2$ pure Calogero Matrix (\ref{matrixop}) $\Omega_{\epsilon=0}$ operator we can introduce the basis of $4$ off-diagonal
hermitian operators. They are given by
\bea
Q_1 &=& \frac{i}{\sqrt 2}(\Sigma_1+\Sigma_3),\label{Q1-free-case}\\
Q_2 &=& i KQ_1 = \frac{1}{\sqrt{2}}(\Sigma_3-\Sigma_1),\label{Q2-free-case}\\
{\widetilde Q}_1 &=& \frac{i}{\sqrt{2}}(\Sigma_2+\Sigma_4),\label{Q1'-free-case}\\
{\widetilde Q}_2 &=& iK {\widetilde Q}_1= \frac{1}{\sqrt{2}}(\Sigma_4-\Sigma_2).\label{Q2'-free-case}
\eea
By construction
\bea\label{qqtildedagger}
Q_i^\dagger =Q_i &,& {\widetilde Q}_i^\dagger = {\widetilde Q}_i\quad\quad (i=1,2).
\eea
Since, for $K$ given by (\ref{klein}), we have
\bea\label{Kanticommutator}
&\{K, Q_i\} =\{K, {\widetilde Q}_i\}=0,&
\eea
the quantum mechanical system defined by $\Omega_{\epsilon=0}$ is ${\cal N}=2$ supersymmetric. Indeed,
\bea\label{N=2susy}
\{Q_i,Q_j\}&=& 2\delta_{ij} {\overline {\bf H}},
\eea
where 
\bea\label{Hcal}
{\overline {\bf H}}&=& {\mathbb I}_2 (-\frac{1}{2}\partial_x^2+\frac{\nu^2}{8x^2})+\frac{\nu}{4x^2}K.
\eea
The following non-vanishing anticommutators are recovered:
\bea\label{N=2sca}
\{Q_i,Q_j\}&=& 2\delta_{ij} {\overline {\bf H}},\nonumber\\
\{{\widetilde Q}_i, {\widetilde Q}_j\} &=& 2\delta_{ij}{\overline {\bf K}},\nonumber\\
\{Q_i, {\widetilde Q}_j\} &=& \delta_{ij} {\overline{\bf D}} +\epsilon_{ij}{\overline{\bf J}},
\eea
where, besides ${\overline{\bf H}}$  given in (\ref{Hcal}), we have
\bea
{\overline {\bf D}} &=& {\mathbb I}_2 (-t\partial_x^2+ix\partial_x+\frac{i}{2}+t\frac{\nu^2}{4x^2})-Kt\frac{\nu}{2x^2},\label{D}\\
{\overline {\bf K}} &=& \frac{1}{2}{\mathbb I}_2 (-t^2\partial_x^2+2itx\partial_x+x^2+it+t^2\frac{\nu^2}{4x^2})-\frac{1}{4}Kt^2\frac{\nu}{x^2},\label{K}\\
{\overline {\bf J}} &=& \frac{1}{2}(K+\nu {\mathbb I}_2).\label{J}
\eea
The four odd (fermionic) operators $Q_i, {\widetilde Q}_i$, together with the four even (bosonic) operators
${\overline {\bf H}}, {\overline {\bf D}}, {\overline{\bf K}}, {\overline {\bf J}}$, close the finite simple Lie superalgebra $osp(2|2)$, as it can be easily checked. By construction $ {\overline{\bf H}}, {\overline{\bf D}}, {\overline{\bf K}}, {\overline{\bf J}}$ commute with $\Omega_{\epsilon=0}$. Therefore, $osp(2|2)$ is an invariant subalgebra of the $2\times 2$ pure Matrix Calogero system.\par
We have indeed, for any generator ${\mathfrak g}\in osp(2|2)$, 
\bea\label{osp22sym}
[{\mathfrak g},\Omega_{\epsilon=0}]&=&0.
\eea
The set of generators $Q_1,{\widetilde Q}_1, {\overline {\bf H}}, {\overline {\bf D}}, {\overline{\bf K}}$ (and, similarly, $Q_2,{\widetilde Q}_2, {\overline{\bf H}}, {\overline{\bf D}}, {\overline {\bf K}}$) close an $osp(1|2)$ superalgebra. $Q_1,{\widetilde Q}_1$, (respectively 
$Q_2,{\widetilde Q}_2$) are odd generators. If, nevertheless, we compute their commutators, we obtain
\bea\label{puredefosc}
&[Q_1,{\widetilde Q}_1]=[Q_2,{\widetilde Q}_2]= \frac{i}{2}({\mathbb I}_2 +\nu K),&
\eea
therefore recovering the ($\nu$-)deformed Heisenberg algebra.\par
For future convenience we also consider the non-hermitian linear combinations
\bea\label{AbarAbardagger}
{\overline A} = i(Q_1-{\widetilde Q}_1), && {\overline A}^\dagger = Q_1+{\widetilde Q}_1,
\eea
which satisfy
\bea\label{AAdagger1}
[{\overline A}, {\overline A}^\dagger]&=& {\mathbb I}_2+\nu K.
\eea

\section{Connection between pure and DFF  Calogero Matrix systems}

It is convenient to normalize the $sl(2)$ root symmetry generators of the pure Calogero system $\Omega_{Cal}$, given in (\ref{omegacal}),  according to
$z_\pm' = \mp i z_\pm$, with $z_\pm$ entering (\ref{purecal}). The corresponding symmetry operators for the pure Calogero Matrix system ${\Omega_{\epsilon=0}}$, given in (\ref{matrixop}), are $z_\pm'{\mathbb I}_2$.\par
It is convenient to redefine $\Omega_{\epsilon=0}$ as
\bea\label{omegaredefinition}
\Omega_{+1} &:= &\Omega_{\epsilon=0}.
\eea
Indeed, a triplet of operators, $\Omega_{\pm 1}, \Omega_0$, carrying a $sl(2)$ representation generated by $z_\pm'{\mathbb I}_2, z_0{\mathbb I}_2$, is encountered. We have
\bea\label{triplet}
\relax [z_-'{\mathbb I}_2,\Omega_{+1}]= \Omega_{0}=2it \Omega_{+1},&&
\relax [z_-'{\mathbb I}_2,\Omega_{0}]= \Omega_{-1} = -2t^2\Omega_{+1}.
\eea
One should note that the commutator $[z_-'{\mathbb I}_2,\Omega_{-1}]=0$ is vanishing, so that no further operator is generated.\par
The operators $\Omega_{\pm 1}, \Omega_0$ close an $sl(2)$ algebra. We have, indeed,
\bea\label{sl2triplet}
\relax [\Omega_0,\Omega_{\pm 1}]= \pm 2 \Omega_{\pm 1},&&
\relax [\Omega_{+1},\Omega_{-1}]= -2\Omega_0.
\eea
The connection between the pure $2\times 2$ matrix Calogero system obtained from (\ref{matrixop}) at $\epsilon=0$  and the $2\times 2$ matrix Calogero-DFF system obtained from (\ref{matrixop}) at $\epsilon=1$  can now be made explicit, by adapting to the present case the construction discussed in \cite{akt}. Under redefinition of the time coordinate and a similarity transformation, the suitably normalized Cartan operator $N \Omega_0$ ($N$ is a constant) is mapped into the DFF $\Omega_{\epsilon=1}$ operator. As a corollary, the symmetry operators of the pure Calogero system ($\epsilon=0$) are mapped into the symmetry operators of the Calogero system with the oscillatorial ($\epsilon=1$) term. \par
The new time variable $\tau$ is introduced from the equation 
\bea\label{newtime}
 t& =& Ce^{2i\tau}, 
\eea
($C$ is here a suitable constant), such that the time-derivative operator entering $N\Omega_0$ can be expressed as
$ 2iNt\partial_t =i\partial_\tau$. This is accomplished by choosing $N=i$, $C=-\frac{i}{2}$.\par
We point out that for our purposes the time $t$ can be complexified and the isomorphism established; a reality condition imposed on the new parameter $\tau$ entering (\ref{newtime}) allows to interpret it as the time for the Hamiltonian with oscillatorial term.\par
The further similarity transformation which maps the symmetry generators of $\Omega_{\epsilon=0}$ into the symmetry generators of $\Omega_{\epsilon =1}$ is given by
\bea\label{simtr}
{\mathfrak g} &\mapsto {\check{\mathfrak g}} = e^{S_2}e^{S_1}{\mathfrak g} e^{-S_1}e^{-S_2},
\eea
where
\bea\label{s1s2simtr}
S_1 = i\tau (x\partial_x+\frac{1}{2}),&&
S_2 = \frac{1}{2}x^2.
\eea
It is rather straightforward to check that, in particular, we get
\bea\label{omegacheck}
i\Omega_0 &\mapsto {\check {i\Omega_0}} = {\mathbb I}_2(i\partial_\tau +\frac{1}{2}\partial_x^2-\frac{1}{8}\frac{\nu^2}{x^2}-\frac{1}{2}x^2) +K\frac{1}{4}\frac{\nu}{x^2}.
\eea
The similarity transformation (\ref{simtr}) and the time redefinition (\ref{newtime}) apply to all $2\times 2$ matrix symmetry generators. We have,
in particular,
\bea\label{sigmaximap}
\Sigma_1 &\mapsto &\Xi_2=e_{12}e^{-i\tau}(\partial_x+\frac{\nu}{2x}-x),\nonumber\\
\Sigma_3&\mapsto&  \Xi_4,\nonumber\\
\Sigma_2 &\mapsto& -\frac{i}{2}\Xi_1,\nonumber\\
\Sigma_4&\mapsto& -\frac{i}{2}\Xi_3.
\eea
The operators in the r.h.s. coincide, for the time variable $\tau$, with the corresponding operators given in (\ref{xis}). It follows, in particular, that $osp(2|2)$ is a symmetry superalgebra of the matrix system with the oscillatorial term. Since the $osp(2|2)$ generators are connected by similarity transformations their (anti)commutation relations are preserved.
\par
It is convenient to define the hermitian generators ${\widehat Q}_1, {\widehat Q}_2, {\widehat Q}_1', {\widehat Q}_2'$ through the positions
\bea\label{qhats}
{\widehat Q}_1 &=& \frac{i}{\sqrt 2}(\Xi_1+\Xi_4),\nonumber\\
{\widehat Q}_2 &=& \frac{1}{\sqrt 2}(\Xi_4-\Xi_1)=iK{\widehat Q}_1,\nonumber\\
{\widehat Q}_1' &=& \frac{i}{\sqrt 2}(\Xi_2+\Xi_3),\nonumber\\
{\widehat Q}_2'&=& \frac{1}{\sqrt 2}(\Xi_3-\Xi_2)=iK{\widehat Q}_1'.
\eea
In the presence of the oscillatorial term we obtain two ${\cal N}=2$ supersymmetric quantum mechanics, since
\bea\label{qhatsusy}
\{ \widehat{Q}_i,\widehat{Q}_j\} &=& 2\delta_{ij} {\widehat H},\nonumber\\
\{ \widehat{Q}_i',\widehat{Q}_j'\} &=& 2\delta_{ij} {\widehat H}'.
\eea
The supersymmetric Hamiltonians are respectively given by
\bea\label{susyhamiltonians}
{\widehat H} &=& {\mathbb I}_2 (-\frac{1}{2}\partial_x^2+\frac{1}{2}x^2+\frac{1}{8}\frac{\nu^2}{x^2}+\frac{\nu}{2})+\frac{1}{2}K (1-\frac{\nu}{2x^2}),\nonumber\\
{\widehat H}' &=& {\mathbb I}_2 (-\frac{1}{2}\partial_x^2+\frac{1}{2}x^2+\frac{1}{8}\frac{\nu^2}{x^2}-\frac{\nu}{2})-\frac{1}{2}K (1+\frac{\nu}{2 x^2}).
\eea
The supersymmetric Hamiltonians correspond to a shift of the original $H_{osc}$ Hamiltonian with oscillatorial term, whose expression is
\bea\label{hamosc}
H_{osc} &=&
 {\mathbb I}_2 (-\frac{1}{2}\partial_x^2+\frac{1}{2}x^2+\frac{1}{8}\frac{\nu^2}{x^2})-\frac{1}{4}K \frac{\nu}{x^2}.
\eea
Indeed, we have
\bea\label{hamhatosc}
{\widehat H} = H_{osc} +\frac{1}{2}(K+{\mathbb I}_2\nu),&&
{\widehat H}' =H_{osc}-\frac{1}{2}(K+ {\mathbb I}_2\nu).
\eea
The shift operators $\pm\frac{1}{2}(K+{\mathbb I}_2\nu)$ commute with $H_{osc}$; we have $\relax [H_{osc},(K+\mathbb I_2\nu)]=0$.\par
An $osp(1|2)$ symmetry superalgebra can be expressed in terms of the operators $A$, $A^\dagger$ and their anticommutators,
where
\bea\label{AAdagger}
A=\frac{i}{\sqrt{2}}(\Xi_1+\Xi_3), &&
A^\dagger = \frac{i}{\sqrt{2}}(\Xi_2+\Xi_4).
\eea
The Hamiltonian $H_{osc}$ is recovered from the anticommutator
\bea\label{AAdaggerosc}
\{A,A^\dagger\}&=& 2H_{osc}.\eea
The remaining two operators ($E_{\pm}$) that, together with $A,A^\dagger, H_{osc}$, close the $osp(1|2)$ superalgebra are given by
\bea\label{newsl2}
\{A,A\}&:=& E_+=e^{2it}[{\mathbb I}_2(-\partial_x^2-2x\partial_x-x^2+\frac{\nu^2}{4x^2}-1)-K\frac{\nu}{2x^2}]\nonumber\\
\{ A^\dagger,A^\dagger\}&:=& E_-=e^{-2it}[{\mathbb I}_2(-\partial_x^2+2x\partial_x-x^2+\frac{\nu^2}{4x^2}+1)-K\frac{\nu}{2x^2}].
\eea
The operators $A, A^\dagger$ induce the deformed Heisenberg algebra since their commutator is given by
\bea\label{AAdgger2}
[A,A^\dagger]&=& {\mathbb I}_2+\nu K.
\eea
We point out that, following \cite{cuhoto}, the Hilbert space can be taken to be given by  a pair of square-integrable functions on the real line. It is then given by a direct sum of lowest weight representations of the $osp(2|2)$ specrtum generating superalgebra, with lowest weights determined by $\nu$. Unlike the $(1+2)$-dimensional theories discussed in \cite{cuhoto} and \cite{anyons}, $\nu$ is not quantized to be an integer or a half-integer value. One should further note that, for $(1+1)$-dimensional $4\times 4$ matrix PDEs, the spectrum generating superalgebra itself, not just its representations, is determined by $\nu$, being given $D(2,1;\alpha)$, with a linear relation between $\alpha$ and $\nu$ (see \cite{cuhoto}).

\section{Deformed Schr\"odinger and higher spin (super)algebras}

We are now in the position to discuss the arising of infinitely generated deformed Schr\"odinger algebra and higher-spin superalgebra as dynamical symmetries of the pure and DFF Matrix Calogero PDEs. Pure and DFF Matrix Calogero dynamical symmetries are isomorphic, since they are related by the time redefinition (\ref{newtime}) and the similarity transformation (\ref{simtr}). It is therefore sufficient to present the results for the pure Matrix Calogero case.\par
The introduction of the deformed Schr\"odinger algebra requires to couple an $sl(2)\oplus u(1)$ invariant subalgebra to a pair of deformed Heisenberg oscillators. Several invariant deformed Schr\"odinger algebras can be recovered as a consequence. Indeed we can pick, e.g., as $sl(2)\oplus u(1)$ subalgebra, either the first-order differential operators $g\cdot{\mathbb I}_2$, with $g$ denoting one of the four operators entering (\ref{purecal}), or the second-order differential operators entering the even sector of the $osp(2|2)$ superalgebra
and recovered from (\ref{Hcal}) and (\ref{D}-\ref{J}). Similarly, up to normalization,  the deformed Heisenberg operators can be given by several possible pairs. We have, e.g.,  
$Q_1, {\widetilde Q}_1$,  $Q_2, {\widetilde Q}_2$ or ${\overline A}, {\overline A}^\dagger$, whose respective deformed Heisenberg commutators are given in formulas  (\ref{puredefosc}) and (\ref{AAdagger1}). 
The choice $\cQ_1=\sqrt{2} {\widetilde Q}_1$, $\cQ_2=-\sqrt{2} Q_1$, so that
\bea
\cQ_1 &=& (e_{12}+e_{21})( it\partial_x +x) +(e_{12}-e_{21})it\frac{\nu}{2x},\nonumber\\ 
\cQ_2&=&- (e_{12}+e_{21})i\partial_x-(e_{12}-e_{21})i\frac{\nu}{2x},
\eea
with
\bea\label{puredef}
\relax [\cQ_1,\cQ_2]&=& i({\mathbb I}_2+\nu K),  \quad (K=e_{11}-e_{22}),
\eea
is particularly convenient. The limit $\nu\rightarrow 0$ exists. At $\nu=0$ (in the absence of the Calogero potential) they coincide, respectively, with the ordinary Galileo boost and space translation multiplied by $e_{12}+e_{21}$ (namely, the operator exchanging the fields entering a $2$-component multiplet). 
The closure, at $\nu=0$, of the ordinary $1+1$-dimensional Schr\"odinger algebra with the extra $sl(2)\oplus u(1)$ first-order differential operators $g\cdot {\mathbb I}_2$ introduced above, immediately  follows.\par
We should recall that, for the original Calogero and de Alfaro-Fubini-Furlan models, the wavefunctions $\Psi(x)$ in \cite{{cal},{dff}} (see also \cite{half}) are defined on the half-line and respect the Dirichlet boundary condition $\Psi(0)=0$ at the origin. It was shown in \cite{mt} and \cite{ftf} that,
in a certain range of the Calogero coupling constant $g$, a more general class of self-adjoint domain for the
one-dimensional Calogero Hamiltonian is allowed. It consists of square-integrable functions defined on the real line (see \cite{mt} and \cite{ftf} for details).  The construction of the Hilbert space for the matrix Calogero and de Alfaro-Fubini-Furlan oscillators induced by a spectrum-generating superalgebra was presented in \cite{cuhoto} and \cite{akt2}. We summarize here the main properties. The Hilbert space is given by an $n$-ple of square-integrable functions on the line. It is determined by a direct sum of the superalgebra lowest weight representations.
 The lowest-weight vectors are defined by imposing the condition $A\Psi_{lw}=0$, $A$ being an annihilation operator (in the present work $A$ is the annihilation operator defined in (\ref{AAdagger})).
A simple inductive proof shows that\\
{\em i}) if $\Psi_{lw}$ is normalized as a $n$-ple of square-integrable functions, all created states $(A^\dagger)^N\Psi_{lw}$ are normalized as $n$-ple of square-integrable functions for any positive integer $N$,\\
{\em ii}) if  $\Psi_{lw}(x) \rightarrow 0$ for  $x\rightarrow 0$, for any positive integer $N$ all created states $(A^\dagger)^N\Psi_{lw}(x)\rightarrow 0$ in the $x\rightarrow 0$ limit (this lowest weight representation can be interpreted to be formed by states belonging to the half-line and satisfying the Dirichlet condition at the origin).\par
The analysis of [12] and [21], applied to the (\ref{matrixop}) operator, proves that:
\\
{\em i}) for $\nu>-1$ the Hilbert space can be selected to be a lowest weight representation with bosonic lowest weight,\\
{\em ii}) for $\nu<1$ the Hilbert space can be selected to be a lowest weight representation with fermionic lowest weight,\\
{\em iii}) in the range $-1<\nu<1$ an alternative choice for the Hilbert space consists in selecting it to be the direct sum of the two (the bosonic and the fermionic) lowest weight representations.\par
At $\nu\neq 0$ the appearance in the r.h.s. of (\ref{puredef}) of the Klein operator $K$ implies that (an infinite tower of) new generators must be introduced to close a Lie algebra. Indeed, for $\alpha=1,2$, we have
$
[\cQ_\alpha, K] =2\cQ_\alpha K
$, so that
$
[\cQ_\alpha , \cQ_\alpha K] =2\cQ_\alpha^2 K$, $
[\cQ_\alpha, \cQ_\alpha^2K] =2\cQ_\alpha^3 K$ and so on.
\par

We can now introduce the associative algebra of Weyl ordered (hermitian) monomials
$$
\cQ_{\a(n)}=\cQ_{\a_1 \a_2 \cdots \a_n}:=\sum_{\a=1,2; \sigma} \frac{1}{n!} 
\, \cQ_{\sigma(\a_1)} \cQ_{\sigma(\a_2)} \cdots \cQ_{\sigma(\a_n)} \,, \quad n=0,1,\dots,
$$
(the $\sigma$'s denote the members of the permutation group acting on $n$ elements), together with their product with the Klein operator $K$. We have, at a fixed value $\nu$,
\be \label{Udef}
Aq(2;\nu):=\{   
(-iK)^b \, \cQ_{\a(n)}\,, \quad n=0,1,\dots,\quad b=1,2 \}.
\ee
The associative algebra $Aq(2;\nu)$ defines the Universal Enveloping Algebra of the deformed Heisenberg algebra \eqref{puredef}. This algebra was first introduced in \cite{vas} in the context of the quantization of observables on a hyperboloid.\par
As a vector space $Aq(2;\nu)$ can be endowed with two different types of brackets:
\par
{\em i}) either ordinary brackets realized by commutators or\par
{\em ii}) ${\mathbb Z}_2$-graded brackets realized by (anti)commutators.\par
In the first case $Aq(2;\nu)$ is an infinite-dimensional Lie algebra (the Jacobi identities being satisfied by construction) that we will denote as 
$q(2;\nu)_{[\,,\,]}$:
\begin{equation}\label{nogradp}
q(2;\nu)_{[\,,\,]}:=\{Aq(2;\nu) \,|\, [\mathfrak{a},\mathfrak{b} ] \,\in \,Aq(2;\nu)\,,\, \forall \, \,  \mathfrak{a},\,\mathfrak{b} \}.
\end{equation}
The infinite-dimensional Lie algebra $q(2;\nu)_{[\,,\,]}$ is another realization of the deformed Schr\"odinger algebra. Since the operators $\cQ_\alpha$ commute with the (\ref{matrixop}) $2\times 2$ matrix operator $\Omega_{\epsilon=0}$, by construction $q(2;\nu)_{[\,,\,]}$ is a dynamical symmetry of the Pure Calogero Matrix PDE; indeed we have
\bea
[(-iK)^b \cQ_{\a_1 \a_2 \cdots \a_n},\Omega_{\epsilon=0}]&=&0.\nonumber
\eea
In the second case $Aq(2;\nu)$ is an infinite-dimensional Lie superalgebra satisfying the graded Jacobi identities and whose consistent ${\mathbb Z}_2$-graded brackets, the (anti)commutators,  are introduced through the position
\begin{equation}
 [\mathfrak{a},\mathfrak{b} \}:=\mathfrak{a}\mathfrak{b}-(-1)^{\epsilon_{\mathfrak{a}}\epsilon_{\mathfrak{b}}} \mathfrak{b} \mathfrak{a}.
\end{equation}
The ${\mathbb Z}_2$-grading of the generators,  $\epsilon_\mathfrak{a}=0,1$, is determined by, respectively, the even or odd power of the $\cQ_\a$'s operators entering any given monomial (the presence of the Klein operator $K$ does not affect the ${\mathbb Z}_2$-grading). We denote the superalgebra, with the given ${\mathbb Z}_2$-graded brackets, as 
\begin{equation}\label{gradp}
q(2;\nu)_{[\,,\,\}}:=\{Aq(2;\nu) \,|\, [\mathfrak{a},\mathfrak{b} \} \,\in \,Aq(2;\nu)\,,\, \forall \, \,  \mathfrak{a},\,\mathfrak{b} \}.
\end{equation}
This $\nu$-dependent class of superalgebras was introduced in \cite{vas} and, together with a Lorentz invariant supertrace formula, used in the construction of higher spin Chern-Simons supergravity (see also \cite{Blencowe:1988gj}).
In \cite{vas} these superalgebras were simply denoted as ``$q(2;\nu)$". We introduced here the ``$[,\}$" label to distinguish them from the ``$[,]$"-labeled deformed Schr\"odinger algebras. We keep, however, the simplified notation when no confusion arises.\par
The infinite-dimensional, $\nu$-dependent, $q(2;\nu)$ superalgebra is a higher-spin superalgebra satisfying the correct spin-statistics connection. In order to understand its relation with higher spin, two main features have to be recalled. The first one consists in the fact that $q(2;\nu)$ contains the finite dimensional Lie superalgebra $ osp(2|2) $ as subalgebra, so that $osp(2|2)\subset q(2;\nu)$; the second feature is due to the fact that $osp(2|2)$ admits a covariant description in terms of three types of indices. They are the vector indices
$\mu,\lambda, \ldots =0,1,2$ labeling a three-dimensional Minkowski space, the (Majorana) spinorial indices $\alpha,\beta,\ldots = 1,2$ labeling the associated real $2$-component spinors and the scalar indices $A,B,\ldots = 1,2$ labeling an internal space.\par
In the covariant notation the $osp(2|2)$ fermions are expressed as $\cQ_\alpha^A$, where one can set
\bea
\cQ_\alpha^1 := \cQ_\alpha ,&& \cQ_\alpha^2 := i K\cQ_\alpha.
\eea
We note that the $\cQ_\alpha^A$'s operators are Hermitian.\par
The even sector is expressed by the $2\times 2$ diagonal operators $J^\mu$ and $R$ (``$R$" stands for the R-symmetry  $u(1)$-generator)  which can be recovered from the saturated (due to the presence of both spinorial $\alpha,\beta$ and internal $A,B$ indices) generalized supersymmetry (see \cite{daf})
\bea\label{gensusy}
\{\cQ_\alpha^A,\cQ_\beta^B\}&=& \delta^{AB}(C\gamma_\mu)_{\alpha\beta}J^\mu+\epsilon^{AB} C_{\alpha\beta}R.
\eea
The Charge Conjugation matrix $C_{\alpha\beta}$ ($C=e_{12}-e_{21}$) and its inverse are used to raise/lower the spinorial indices; the diagonal metric $\eta_{\mu\nu}$ ($\eta_{\mu\nu}=diag(-1,+1,+1)$) and its inverse raise/lower the vector indices; the three gamma matrices ${(\gamma_\mu)_\alpha}^\beta$ ($\gamma_0=e_{12}-e_{21}$, $\gamma_1=e_{12}+e_{21}$, $\gamma_2=e_{11}-e_{22}$) satisfy the composition law of the split-quaternions given by
$\gamma_\mu\gamma_\nu = \eta_{\mu\nu}{\mathbb I}_2 +\epsilon_{\mu\nu\lambda}\gamma^\lambda$, with the totally antisymmetric Levi-Civita pseudo-tensor $\epsilon_{\mu\nu\lambda}$ normalized so that $\epsilon_{012}=1$; by definition $(C\gamma_\mu)_{\alpha\beta} = C_{\alpha\alpha'} {{(\gamma_\mu)}_\beta}^{\alpha'}$. It is worth pointing out that the Charge Conjugation matrix and the gamma matrices act on the spinorial indices of the algebra and not on the $2$-component wave function of the original PDE. The normalization of the antisymmetric tensor $\epsilon^{AB}$ is $\epsilon^{12} = 1$.\par
Straightforward computations show that the $2\times 2$ matrix differential operators $J^\mu$, $R$ are connected with the ${\overline {\bf H}}, {\overline{\bf D}}, {\overline{\bf K}},{\overline {\bf J}}$ operators of formulas (\ref{Hcal}) and (\ref{D}-\ref{J}) through the positions
\bea
&J^0=-2({\overline{\bf K}}+{\overline{\bf H}}), \quad J^1=2({\overline{\bf K}}-{\overline{\bf H}}),\quad J^2=2{\overline{\bf D}},\quad R=2{\overline{\bf J}}.
&
\eea
The closure of the $osp(2|2)$ superalgebra is guaranteed by the non-vanishing commutators
\bea
\relax [J_\mu,J_\nu]&=& 4i \epsilon_{\mu\nu\lambda}J^\lambda,\nonumber\\
\relax[J_\mu, \cQ_\alpha^A]&=& 2i {(\gamma_\mu)_\alpha}^\beta \cQ_\beta^A,\nonumber\\
\relax [R, \cQ_\alpha^A] &=& -2i {S^A}_B \cQ_\alpha^B.
\eea
In the last equation the matrix ${S^A}_B$ is given by $S=e_{12}-e_{21}$.\par
We point out that the construction of the above superalgebra in term of a Klein operator was discussed in \cite{ply}.
Due to the presence of the covariant $osp(2|2)$ subalgebra, the remaining generators of $q(2;\nu)$ are accommodated into higher spin representations.\par
As mentioned before an associative algebra, called ${\widehat{Aq}(2;\nu)}$, of invariant operators for the DFF Matrix Calogero PDE is recovered from ${{Aq}(2;\nu)}$ by applying the time redefinition 
(\ref{newtime}) and the similarity transformation (\ref{simtr}).  From ${\widehat{Aq}(2;\nu)}$ one recovers the deformed Schr\"odinger algebra ${\widehat{q}_{[,]}(2;\nu)}$ and the higher spin superalgebra ${\widehat{q}_{[,\}}(2;\nu)}$. They are isomorphic realizations of ${{q}_{[,]}(2;\nu)}$ and ${\widehat{q}_{[,\}}(2;\nu)}$, respectively.

\section{Conclusions}

As we have seen the Matrix Calogero models under consideration possess, in particular, a deformed Schr\"odinger algebra and the higher spin superalgebra  $q_{[,\}}(2;\nu)$  as dynamical  symmetries.
 The deformed Schr\"odinger algebra is obtained by computing, similarly to \cite{nie1}, the class of linear operators which leave invariant the space of solutions of the given PDEs. A whole universal enveloping algebra, $Aq(2;\nu)$, is generated by requiring the closure of the Lie algebra induced by the commutators. When the Calogero potential coupling parameter $\nu$ takes the $0$ value, the infinite dimensional algebra can be truncated to the standard Schr\"odinger algebra.\par
We further observed that the matrix Calogero models and (at $\nu=0$) the free particle multiplet, exhibit (relativistic) higher spin symmetries. These results suggest that, extending the conjecture of \cite{Valenzuela2016}, the non-relativistic Calogero system studied so far may be dual, in a certain regime and in a certain asymptotic non-relativistic geometry (see \cite{Duval:2008jg}), to the vacuum of the (bulk) Vasiliev's higher spin gravity. As a matter of fact, the $\nu$-parameter is related to the vacuum expectation value of the scalar field in the Vasiliev-Prokuskin higher spin gravity \cite{Prokushkin1999} in $2+1$ dimensions. 
\par
We should mention that another class of $(1+1)$-dimensional non-relativistic PDEs, unlike the Matrix Calogero models here discussed, inherently  contain higher spin operators \cite{akt0,akt}.  These PDEs are invariant under the centrally extended ${\ell}$-Conformal Galilei algebras, with ${\ell}=\frac{1}{2}+{\mathbb N_0}\geq\frac{3}{2}$. In these models the higher-spin generators span a spin-${\ell}$ representation of an $sl(2)$ subalgebra constructed from ${\ell}+\frac{1}{2}$ Heisenberg algebras (${\ell}=\frac{1}{2}$ for the free particle in one dimension). An open problem consists in understanding whether, for these systems, a
deformation of the
centrally extended ${\ell}$-Conformal Galilei algebra, induced by matrix-Calogero potentials, is allowed.\par 
We expect, more generally, that systems  possessing conformal Galileo invariance could be related to higher spin gravities. This type of non-relativistic dualities, if they indeed exist, would complement the dualities conjectured in \cite{Sezgin:2002rt,Klebanov:2002ja} (see also \cite{Maldacena:2008wh}).
~
\\ {~}~
\par {\Large{\bf Acknowledgments}}
{}~\par{}~\par

F.T. received support from CNPq (PQ Grant No. 306333/2013-9). He thanks the Universidad San Sebasti\'{a}n of Valdivia, where this work was elaborated, for hospitality. M.V. thanks as co-investigator the projects Fondecyt Regular No 1140296 and Conicyt DPI 20140115.


\end{document}